# Social software and strategy

Special issue editors

Stefan Haefliger, Eric Monteiro, Dominique Foray, and Georg von Krogh

ETH Zurich, NTNU, EPFL

June, 2011

## *Introduction*

Aligning interests, motivating contributions to knowledge work, and giving direction to multiple business units and market initiatives represent daily challenges facing the strategist in most companies. The diversity of contexts within an organization has thus led critical management thinkers to suggest that in the presence of multiple initiatives and discourses, the introduction of a new technology has multiple unintended consequences for organization. New technology is regularly subject to power struggles, conflicting goals, and discrepant events (Markus, 1983; Barley, 1986; Orlikowski, 1992; Ciborra, 1996; Leonardi, 2008) which impact on how strategies are shaped within organizations.

Information technology, such as social software, may affect the interaction patterns between organizational members, create new opportunities for knowledge and information sharing (von Krogh, 2002) or unfold the disruptive and possibly change-inducing potential of so called "informational capabilities" (Leonardi, 2007).

Informational capabilities refer to an information technology's potential to alter the storage, transmission, and creation of information in an organization. Information and communication technology (ICT) differs in this respect from other technologies adopted by organizations. Despite the broad application of ICT and its potential implications for company performance (Powell and DentMicallef, 1997; Tippins and Sohi, 2003; Ho et al., 2011), strategy scholars seldom include the properties of ICT in their theorizing on strategic thinking, firm growth and its boundaries, or the strategies for creating ICT infrastructure (Leidner et al., 2010; Yoo et al., 2010).

The term "social software" grew out of the notion of groupware and computer-supported collaborative work (CSCW) and had been attributed to Clay Shirky denoting software that supports group interaction (Allen, 2004; Shirky, 2005). Designed to facilitate individual creativity combined with community building, groupware and social software has led, in its 1.0 incarnation, to novel and significant insights in academic fields ranging from technological innovation, organization behavior, management and organization theory, to strategy (Sproull and Kiesler, 1986; Sawhney and Prandelli, 2000; Lee and Cole, 2003; von Krogh and von Hippel, 2006). Today, social software, frequently annotated with web or enterprise 2.0, also receives a lot of interest from managers due to commercial use, increased network functionality, massive mobilization of users in some cases, and growing infrastructure capabilities such as multi-media streaming online[1]. The more than enthusiastic reception of LinkedIn by investors during their initial public offering in May 2011 suggests that the commercial promise of a business model involving large numbers of users connected through social software inspires investors: secondary market valuations of companies such as Facebook, Groupon, or Twitter are interpreted along similar lines or else discounted as signs of market participants' exuberance.

---

[1] For the original definition of Web 2.0 see O'Reilly, 2005. Wirtz, Schilke and Ullrich (2010: 276) characterize Web 2.0 with four factors: social networking, interaction orientation, personalization/customization, and user-added value.

Social software affects the interaction between employees within and individuals outside the firm, such as members of user communities or customers. In many industries, users of technology, frequently organized in communities, are known to innovate independent of manufacturers (von Hippel, 1988; 2007), and consumers have successfully contributed to innovation and product development organized by firms (Füller et al., 2010; Franke et al., 2010). A perspective that privileges firm-internal matters, as the strategy field has often tended to adopt (Minzberg, 1978), risks overlooking the increasingly powerful and important position that individuals outside the firm hold, particularly when organized in communities (Fredberg, 2010; Dahlander and Wallin, 2006). Here, users and customers set up the governance structures for their communities independent of firms (Markus, 2007; O'Mahony and Ferraro, 2007), often voice criticism towards firms and their products (Kaplan and Haenlein, 2010; Kozinets and Handelman, 2004; Muniz and O'Guinn, 2001), or develop rival products in existing markets (Casadesus-Masanell and Ghemawhat, 2006; Young and Rohm, 1999). Hence, in terms of strategic analysis, users and consumers can be "suppliers," "competitors," or "providers of substitutes," as shown in the examples of widely used Open Source software programs like Apache servers or the GNU Linux operating system.

Social software enables the activities of people inside and outside companies to actively shape strategies. The papers in this special issue offer novel, strategic insights on this subject. Social software thus holds an intriguing potential. As consistently demonstrated over the last couple of decades, the actual realization of this potential is anything but straightforward. The special issue unites authors who gained deep insights into the workings of user communities, their technologies, and the potential that collaborations between firms and communities harbor for strategy in terms of value creation and -appropriation. This introduction article contributes to an overall positioning of the papers drawing on the literatures from fields such as strategy, information and communication technology (ICT), technology studies, and innovation.

In so doing, we also develop a research agenda on social software that hopefully will inspire strategy scholars to continue work in this important area.

Social software creates platforms for self-expression (Schau and Gilly, 2003) and direct interaction between individuals and thus facilitates rapid and often spontaneous community building (Culnan et al., 2010). Social software also enables interaction among consumers and users online, their efforts in product development (Füller et al., 2010), the mutual rating of ideas and comments among themselves (Reichen et al., 2009), and provides online community members with a basic infrastructure for their work (Lee and Cole, 2003; Ren, Kraut and Kiesler, 2007; Kohler et al., 2011). Thus, social software becomes an exciting topic for strategy practitioners and scholars. If social software supports management in harnessing the creative output of individuals inside and outside the firm the deployment and diffusion of such technology may hold a considerable business potential. The video game industry, for example, experiences rapid growth thanks to social software platforms (such as Facebook with more than 500m registered users) that serve as alternatives to consoles and enable online gaming that involve competition among friends connected through such platforms[2].

At the same time, the "harnessing" metaphor often employed by companies interacting with communities, may be problematic. It might very well suggest too much influence by the firms. The connotations of control embedded in the metaphor need to be supplemented or even substituted by stronger aspects of cultivation and facilitation. Evidence for successful interaction between firms and user communities is scarce (e.g. Stam, 2009; Stuermer et al., 2009) despite frequently high investments by firms in such collaboration (Dahlander and Wallin, 2006). IBM, for example, invested significant resources into the public development of their Eclipse software development platform for a duration of five years, before development by outside software users outweighed IBM's own development efforts (Spaeth et al., 2010). Three implications result from the use of social software that favors a broader view of collaboration, extending beyond the company. First, consumers and users can assume several roles of strategic importance for the company beyond the role as customers of products and services. Most notably, they may supply ideas for product development (see Fuchs and Schreier, 2011), and may offer competing products as in the case of Open Source software (see von Hippel, 2007). Second, social software shares with all information technology the capacity to change organizations in unpredictable ways, because it directly alters the way and the location where information is stored, shared, and created (Kling and Scacchi, 1982; Markus, 1983; Leonardi, 2007; Leonardi, 2008). The fact that most "outside members" of social software platforms are unknown to the firm makes it even harder to foresee how ICT will change the organization. Third, consumers and users rely on social software to organize in online communities that may or may not be supported by companies, and "develop a life of their own (Wiertz and Ruyter, 2007:390)." Understanding which interventions by the company will be perceived as beneficial or obtrusive is key to building lasting relationships with members of such platforms (Jeppesen and Fredriksen, 2006). In the software industry, IBM received credit for their efforts to support the Open Source community, whereas Sun Microsystems (now Oracle) was widely criticized for their hesitation to release the source code for Java, the cross-platform programming language (West and Gallagher, 2006; Vaughn-Nichols, 2009).

---

[2] For example, Zynga, the producer of games such as FarmVille, is valued at 10b USD and thus significantly higher than, e.g., game industry incumbent Electronic Arts (7.5b USD) according to the Wall Street Journal (Feb 19, 2011). And while 90% of all users play for free, 10% of the 250m users reliably pay small amounts for in-game assets and enhancements.

    These three implications serve in building a framework that we present in the next section. The framework organizes research on social software taking perspectives from both inside and outside companies, and we use it to locate the contributions to this special issue in terms of strategy, technology, and community. In the third section we use the framework to highlight open issues for strategy research building on the contributions by authors in this issue.

### *Towards a framework*

Social software has been in use in firms for a number of years but standard or best practice ways of applying social software are not visible yet. Managers and researchers alike still struggle with questions, such as why and how to interpret social software, what are shared perceptions, how to appropriate potential business value, when to enact work practices involving social software, and where to align it with other business processes. For example, the usefulness of social software as an internal communication device is up for debate (see Denyer et al., in this issue). Moreover, the perception of Facebook as a platform to do business is changing the gaming industry. Finally, aligning internal software development efforts with external community development creates new challenges for design science (von Krogh and Haefliger, 2010).

The role of information and communication technology in organizations has been the focus in decades of research (Markus and Robey, 1988; Leonardi and Barley, 2010). Early organizational theorists considered technology to have a unidirectional impact on organizations forcing management to change some aspect of the organization according to the contingencies inherent in the material features of the technology (Perrow, 1967). This perspective was later challenged by social constructivists who focused instead on the members of the organization who responded to the technology's constraints and to each others' use of the technology (for a review, see Leonardi and Barley, 2010) as well as on the material features of the technology and its role as an actor in organizations (Orlikowski and Scott, 2008; Wagner et al., 2009). This highlights the inherently unintended consequences of technology which undermines overly instrumental 'deployment' of technology (Rolland and Monteiro 2007). It is this active role of ICT as a mediator between individuals in organizations and between intended and completed action that complicate strategic deployment and adoption of ICT in firms, even more so when individuals outside and frequently unknown to the company play an important role.

The study of social software for the implications it holds for strategy needs to accelerate since many companies are far ahead with experiments connecting individuals inside and outside the company. At the same time, strategy research should broaden the narrow perspective of authoritative decisions about technology adoption that might miss the influential role technology plays outside the direct control of management. With this special issue and this framework we attempt to follow both suggestions. More specifically, we suggest issues for future research to build on a balanced perspective that takes into account what management can and should influence combined with an appreciation of consumers' and users' work outside the firm. Two observations about social software may help strategy scholars understand the connection between the perspectives from inside and from outside the firm. First, social software shapes the behavior of individuals during evaluation, adoption, early use, and adaptation. Second, and closely connected to the first point, it enables individuals inside and outside the firm to appropriate features of the technology in ways unintended by management or the technology's designers (Markus and Silver, 2008; Poole and DeSanctis, 2004; DeSanctis and Poole, 1994). The case of LEGO (Hienerth et al. in this issue) shows how the adoption of social software enabled a business model where customers co-create new products and commercialize them on the LEGO platform. In the process LEGO had to overcome significant organizational and psychological barriers including the fear of losing control. Stuermer et al.'s (2009) study of Nokia's development of the Open Source Maemo platform reached a similar conclusion.

There are a number of recent contributions in strategy and organization theory that have addressed issues involving social software in the domains of strategy, technology, and community. Table 1 presents the proposed framework and the research published in this special issue. The table distinguishes work that takes a view from inside the company, from studies that have focused on similar issues outside the firm. Rather than aiming at an extensive review of this literature, below we focus on distinctive characteristics of the two views that are relevant for understanding the strategic dynamics social software can help generate. First, research that bridges the two perspectives has emerged in the management of innovation (Rossi Lamastra, 2009; Fuchs and Schreier, 2010; Capra et al., 2011), but less so in strategy, management and organization theory (O'Mahony and Bechky, 2008). The articles united in this issue take on the task of bridging insights that emerge from studying managerial intentions as well as user and consumer behavior inside and outside the firm. Studies limited to a view from outside the firm may bear little business relevance and studies limited to a view form inside the firm may ignore activities by (sometimes unknown) outsiders with significant potential impact on strategy and new business models.

Insert Table 1 here

**Strategy**

Already thirteen years ago, Jeffrey Sampler (1998: 349) proposed that the availability of critical information for the same market defines industry boundaries, rather than what strategy scholars' considered an industry; firms delivering comparable or similar outputs. Thanks to the free exchange of information over the Internet and the access to social software applications, information may be available to individuals inside or outside of companies, that assumes a strategic importance for a market. Such information may include insights into customer preferences as they are shared in social networks, ideas for new products and services, or information about available substitutes. On the one hand, research on consumer and user communities in the areas of marketing and user innovation focused on a view outside the firm, and showed that consumers and users build communities and organize to achieve specific goals (Muniz and O'Guinn, 2001; Moon and Sproull, 2001; O'Mahony, 2003; Wiertz and Ruyter, 2007). On the other hand, strategy research looked inside the firm to approach the question of how firms can make use of consumer and user communities in the creation and appropriation of value. The results from implementing social software here, however, tend to be focused on specific positive cases or they provide inconclusive results (da Cunha and Orlikowski, 2008, see also Denyer et al., this issue). Moreover, the growing literature on open innovation tends to confine the exchange with external parties, such as suppliers or users, to identifiable and manageable knowledge (Chesbrough, 2003), such as research papers, information about patents or instruments, or to one-directional search in a space of technological opportunities (Laursen and Salter, 2006; Jeppesen and Lakhani, 2010). An exception can be found in new product development, where firms have started to successfully empower customers to interact among themselves (Fuchs and Schreier, 2010). Using social software in its simpler (1.0) versions, firms have successfully appropriated value from implementing strategies targeted at collaborating with open source software development communities (Henkel, 2006; Dahlander, 2007). However, so far only Stam (2009) conclusively linked collaboration with open source communities with innovation performance. Frequently, the argument in the literature is that devoting resources to collaboration strategies with user communities would only be beneficial, if their contribution pays off for the firm (von Hippel and von Krogh, 2003; Dahlander and Wallin, 2006).

    A view from outside the company suggests that social software generates value for individuals because it facilitates interaction with and learning from other consumers and users, helps to build shared identity, and enables joint creation and shaping of technology for own use (Lakhani and von Hippel, 2003; Kuk, 2006, Hertel et al., 2003; Spaeth et al., 2008). Initially, so-called "commons based peer production" relied on users

who connected and exchanged information through a set of rather simple social software tools such as email lists, Internet relay chat, and message fora (Benkler, 2002; Lee and Cole, 2003). While the technology became more sophisticated (2.0) the communities spread and grew: individuals perceived value in collaboration and continued to exchange information in online communities (Ren et al., 2007). Meanwhile, much of the value generated is free and publically available (including posts in online fora and Open Source software) and consumers and users take measures of protecting this value (O'Mahony, 2003), such as non-profit incorporation, social norms of collaboration, and legal refinement (e.g. licenses under which Open Source software are made available). Licenses are designed to keep access to information and technology as open as desirable or possible, and they include creative commons- and various Free and Open Source software licenses (Lerner and Tirole, 2005). Furthermore, users turn entrepreneurs by learning from industry experts and recruiting through a network enabled by social software (Haefliger et al., 2010).

The first contribution to this special issue by Burger-Helmchen and Cohendet provides guidelines and examples of how companies in the video game industry foster special relationships with community members outside firm boundaries, in order to gain insights and creative output from collaborating with closely bound and loyal customers. The authors classify communities and their members in order to better understand appropriate firm action to improve the relationships with communities because co-creation of value is a fragile process that depends on motivation and mutual trust. The second contribution by Hienerth, Keinz and Lettl explores the characteristics of user-centered business models building on well-known cases such as LEGO and IBM. A core contribution to strategy lies in identifying successful strategies for integrating users into core business processes and for overcoming internal resistance. The authors elaborate how these processes enable the appropriation of value as part of a new user-centered business model.

**Technology**

Information and communication technology is a tool for management but, at the same time, it harbors a deeply disruptive potential for organizational change that should not be underestimated. Users inside and outside companies attribute meaning to the functionality offered by a technology which can alter the identity of a technological artifact such as a search platform or a discussion forum (Faulkner and Runde, 2009), change work practices such as information seeking (Leonardi, 2007), or, as put by Shirky (2003), result in a "runtime effect" of ICT. In analogy to software certain characteristics of the program become apparent as a runtime effect: only after an ICT system is installed and used, we can discern its real impact on the organization. Bridging economics to management, Brynjolfsson and co-authors (2009) discuss evidence on key aspects of how firms have transformed themselves by combining IT with changes in work practices, strategy and products and services; they have transformed the firm, supplier relations, and the customer relationships. Both case studies and econometric work point to organizational complements such as new business processes, new skills, and new organizational structures as major drivers of the contribution of information technology. The management literature often uses suggestive wording about "harnessing" and "utilizing users'" creative thinking. Frequently, management scholars have adopted a perspective on ICT as a tool for gaining access to users' creative output (e.g. Füller et al., 2010). Less often, authors have focused on the biases and novel forces social software introduces when mediating and organizing work. For example, Dellarocas and Wood (2008) demonstrated the impact of online trading platforms on the interaction between users in a way that resulted in massively overstated user satisfaction. In a study on market research, social software is shown to have the potential to generate insights due to the links it enables between consumers, and their commenting and rating behavior that introduces quality judgments and points to trends (Cooke and Bukley, 2008). In these examples, social software mediates between individuals' activities and collective outcome in ways that are limiting or enabling.

Users perceive social software as a tool for creative expression and identity building online (Schau and Gilly, 2003; Muniz and O'Guinn, 2001). Visibility and peer recognition motivate consumers and users to share personal experiences with products and companies, and even lead to the development of sub-cultures with their specific vocabularies, creative expression, and behavior (Kozinets, 2002). Conversely, users are shaped by social software, the architecture of digital artifacts, and the specific practices of collaboration that surround and build these artifacts. According to Baldwin and Clark (2006) the architecture of a software can be understood in terms of "option value" where collaboration and contributions to its development are guided by the user's perceived rewards in terms of progress and recognition. For example, users are known to self-select into tasks for a collaborative project in Open Source software development (Yamauchi et al., 2000) and the specialization of labor in projects follows the logic of an evolving and growing technology implementation (von Krogh et al., 2003). Hence, there are strong linkages between the architecture of an ICT system and the behavior of users, including where and to what part of the technology they choose to contribute, how they collaborate and communicate, or even when and where they choose to free-ride on what other users provide.

Contributors to this special issue approach the technology of social software from a strategic perspective and with particular attention to the motivations and reservations of individuals, inside and outside the firm. The article by Denyer, Parry, and Flowers documents the effort to deploy social software within a large telecommunications company. Their story is one of disappointment relative to the high-flying promises of openness and participation. The authors find that the solution implemented did not achieve positive outcomes relative to more traditional methods of communication. The authors offer valuable insights into political processes, such as monitoring and self-promotion, that may have contributed to the dismal reception of the new technology. They show that the problems discovered do not lie with the technology but with the behavior of the users who need to find a delicate balance of power between leaders and employees of the organization. The study by Frey, Haag, and Lüthje studies a search platform for innovative ideas. Social software takes the form of a mediator between individual contributors and firms performing broadcast search on the platform. The authors suggest that deploying such a platform leads to more substantial contributions, when it succeeds in attracting intrinsically motivated individuals with diverse knowledge backgrounds. Both these contributions engage with technology and lead the authors to caution strategists in being too ambitious towards social software and urging them to take the perspective and motivation of the user seriously.

**Community**

Social software is an integral part of the formation of online communities; it enables individuals to interact and socialize, who may not be previously linked. Part of the strategist's fascination with social software stems from the possibility to access a pool of voluntary contributors to strategy, products, services, and business models, who are qualified, motivated, and productive. Realizing this potential demands influence, which is not easy to gain. Dan Frye, Vice President of Open Source software at IBM, commented on IBM's work with the Eclipse community: "There is nothing that we can do to control individuals or communities, and if you try, you make things worse. What you need is influence. It goes back to the most important lesson, which is to give back to the community and develop expertise. You'll find that if your developers are working with a community, that over time they'll develop influence and that influence will allow you to get things done." (Quoted in Kerner, 2010.)

The question is what strategic actions towards facilitating community interaction are possible, for whom and at what stage (Thompson 2005)? Leadership in an online community is fragile because gaining influence takes years of commitment and investment (Spaeth et al., 2010) and since the involvement of companies may change community members's motivation (Shah, 2006; Stewart et al., 2006). For example, companies must decide whether to found a community or to sponsor an existing community (West and O'Mahnoy, 2005). Both options involve trade-offs with regards to control, influence and the costs. Users may look beyond the deployment of social software to consider joining existing social networks that span beyond and across comapnies. Thus, it becomes important to understand the governance structures of online communities in order to appreciate the differences and potential risks when applying established leadership practices from a corporate context (Markus, 2007; O'Mahony and Ferraro, 2007). Leadership in user communities is thought to emerge from a meritocracy where technical achievement and boundary spanning is rewarded with power (Fleming and Waguespack, 2007). O'Mahony and Ferraro (2007) drew a refined picture by showing that technical skill alone does not lead to powerful positions: social skills of mediation and negotiation among community members predict future leaders more reliably (see also Fleming and Waguespack, 2007; Collier et al., 2010).

Community boundaries form around individuals, frequently volunteers, who "interact over time around a shared purpose, interest, or need (Ren et al., 2007: 378)." Beyond a general understanding of the risks in social software, such as "knowledge leakage" (e.g. Hustad and Teigland, 2008), the current strategy literature offers little guidance for firms how to manage community boundaries. Central questions involve the selection, joining, and adherence to norms in existing communities and how this related to staffing, task allocation, or business process involvement. Professional communities play an important role in the early stages of alliance formation (Rosenkopf et al., 2001) but comparable research on online communities is largely absent. A potential recourse for strategy scholars may be the literature on virtual teams that emphasizes the role of facilitators and cautions about the risk inherent in spanning different cultural contexts (e.g. Pauleen and Yoong, 2001; Martins et al., 2004). This risk may grow when linking corporate and non-corporate contexts.

The practice of setting up mechanisms to protect intellectual property reveals that users are similarly concerned about losing control over their work as companies are (O'Mahony, 2003). However, users often operate in a context of private-collective innovation outside corporate hierarchy and without labor contracts that regulate their contributions to the community or company (von Hippel and von Krogh, 2003). Due to this constellation, researchers have devoted comparatively more attention to the motivation of users with respect to community boundaries than to leadership issues involving the firm. The motivation to contribute to the community seems to be affected by whether companies are involved and sponsor the community or not (Shah, 2006; Stewart et al., 2006), and the extent to which the firm explicitly credits and recognizes contributions by users (Jeppesen and Fredriksen, 2006).

Two contributions in this special issue deal explicitly with the interactions between firm and community. Sutanto, Tan, Battistini, and Phang test a model of emergent leadership in a setting where users interact and develop network ties. The model predicts perceived leadership from interaction patterns of users and may, thus, provide strategists with specific insights and potential levers on companies interacting with users communities. The work by Jarvenpaa and Lang focuses on community boundaries by taking a holistic perspective on companies and users forming communities. They also discuss interdependencies and negotiations necessary when managing what the authors call the "generative capacity" of online communities, that is their ability to rejuvenate, reconfigure, reframe, and revolutionize around the members' shared purpose.

This special issue on social software assembles works that span the perspectives from inside the firm to outside the company, by studying topics of relevance to strategy and putting attention on role of consumers and users. In the following, we build on these contributions by formulating an agenda for research going forward

## *Open issues for strategy research*

There are several open issues regarding social software that deserve the attention of strategy scholars. Many of them start with a practical appreciation of the business implications of this technology; Practical technologies for recruiters may help human resource management refine their frameworks for talent management and succession planning; New ways of storing, accessing, and locating patient data may bring about not only personalized medicine but also changes in health management systems; Best practices of compensating resourceful users boost new product development initiatives.

Generally, strategic management is concerned with issues such as firm survival, the allocation of resources across business units, or the creation of novel business models. In what other areas than those treated in this special issues does social software impact on these questions and how? To begin exploring such issues, it is worthwhile to consider the larger ramifications of social software. While refined definitions of social software may moderate or limit disruptive effects to specific business processes, the logic builds on what Leonardi (2007) called informational capabilities of information technology. Organizations need to grapple with fundamentally indeterminate effects when introducing social software at many levels. The idea of a "runtime effect" of social software (Shirky 2003) refers in analogy to the role the system environment plays in the execution of a software program. Adoption, use, and adaptation of a new technology, such as social software, provide contexts in which organizational actors define what a technology means and can do for them before and during action (see Leonardi and Barley, 2010 for a review). Hence, the "management" of social software becomes an ongoing task that incorporates the user's role and adapts strategy according to negotiation and structuring of work. Against this background, we develop a series of questions for strategy research that pays particular attention to value creation and appropriation, the role of technology both as tool and mediator between managers and users, and the role that management may play in communities as leaders and in shaping boundaries. Table 2 contains the questions the framework evokes.

Insert Table 2 here

**Co-creation and appropriation of value**

The creation of economic value that involves consumers and users connected through social software may depend on organizational structures that support this work. The individuals may or may not be members of the same organization. Yet, the new links between individuals, the exchange of information, and the potential to adhere to the norms of such a network may generate opportunities for knowledge sharing and joint work inside an organization that could be very valuable, or even disruptive to existing ways of creating value. The first question regarding organizational structure touches upon fundamental issues in strategy: which parts of hierarchy remain intact and which ones may change? How may social software impact on formal and informal organization and their interaction? How are decision rights allocated amongst members in business processes with open networks and free flows of information? Who gets authority to interact with external users? What are the "hidden costs" of changes in organization structure?

The issue of value creation has an important time component in that co-creation between firms and outside consumers and users involve building trust, providing mutual support, and bearing joint questioning. If social software is to grant access to members from outside the company, the meaning of a "common purpose" may change. Some consumers show extraordinary loyalty to brands and products over a long period of time. Creating a shared purpose relating to a brand or a product could be a productive way of activating value co-creation. This may be costly and time-consuming and the question as to what supports mutual buy-in remains open to research. In their seminal study, Jeppesen and Fredriksen (2006) showed that explicit recognition of outside contributions had a positive impact on value creation.

Value appropriation requires relatively exclusive access to an asset or complementary assets that allow for products or services to be derived. The growth of business models that contain some "free" elements, and using advertisement to collect revenue (McGrath, 2010), indicates that appropriating value from user-generated content may become easier. However, the creative commons family of licenses may lead to the growth of the number of domains where appropriation of others' works becomes less straightforward, and companies can no longer count on unsuspecting users who, sometimes ignorantly, pass on the rights to their intellectual property (von Hippel, 1988). With the growing awareness of intellectual property infringements, we also expect more public awareness of ownership. The creative commons movement actively educates users about their rights and advocates that they make a conscious choice about how to license creative work[3]. In software, Open Source software licenses limit the possibility of users and firms to appropriatiate the rights to software components for re-sale. The important works by Henkel (2006), as well as Dahlander and Magnusson (2007; 2008), have classified a series of strategic approaches to this difficulty encountered by software companies.

Future research may uncover generic patterns in business models that take advantage of assets co-created with consumers and users. The work by Hienerth and colleagues in this issue takes a major step in that direction. When following McGrath (2010) who suggested that successful business model innovations are discovery driven, the issue of "runtime effects" of ICT may even prove to be an advantage for firms that experiment with technology such as social software. Once deployed and subject to adaptation, social software platforms may evolve in unpredictable directions. McGrath (2010: 254) points out that business model experimentation takes place across and within companies. Thus, the use of platforms such as Facebook may alter information flows across and within firms leading to new opportunities for products and services. Consider Zynga, the producer of online games: friends already connected via the social software platform (Facebook) may compete against each other in online games for free or acquire certain in-game assets for improved performance, and so on. Cross promotion activities among games published by Zynga may retain customers or introduce further products and services as the user base grows. The notion of business model portfolios (Sabatier, 2010) could be a promising starting point for scholars who want to theorize about complementary strategies for value appropriation using social software.

---

[3] For further information about creative commons turn to: http://creativecommons.org/about/

**Contextualizing social software**

From the perspective of changing opportunity structures and informational capabilities, social software is both a tool and mediator in organizational processes. Maintaining balance and achieving specific goals from a managerial point of view entails paying attention to four factors: context, power, ethics, and trust. Regarding future research, all of these factors deserve more explicit attention in order to support strategists. First, social software is applied to a specific organizational context or business process. There is a choice of maintaining and supporting an existing context or accommodating work involving social software. Are certain business processes more amenable than others to work practices involving social software? Do contingencies play a role, such as hierarchical information barriers, openness to new organization members, privacy issues, or prior communication patterns? Furthermore, does adaptation of the social software change its reception in the organization? Does specific type of work go better with social software than others? Can strategy processes be opened to outside participants through social software? Strategists should not forget that ICT can be heavily customized or designed in-house. Hypothesizing about contingencies and adapting after adopting technology, may pave the way towards creating a favorable context for using social software in a way that can be perceived as successful by both users and management. Case studies of more or less successful implementations of social software along the lines of the contributions in this special issue may help to identify additional context factors.

Second, power struggles play an important role in a number of areas of technology management, from the viewpoint of institutional theory (Hargrave and Van de Ven, 2008) as well as from an organizational perspective (Leonardi and Barley, 2010). Proponents of specific technologies form networks (Garud et al., 2002) or change institutions (Hargrave and Van de Ven, 2008) by leveraging and applying legitimacy and framing strategies to supersede opponents. On a micro scale, actors within one organization or community may dominate others in defining modes of use and work practices involving social software. Leonardi and Barley (2010) suggest that because the construction of meaning and organizational change occurs at multiple levels and phases of ICT implementation and because its outcome is indeterminate both the activities of humans and the material features of the technology matter for the outcome of organizational change. Power struggles may well determine the outcome of strategic initiatives and challenge strategic management in terms of organizational justice and fairness. A pertinent question in this regard is who is allowed to access information on social software platforms, and for what purpose (for a review see Colquitt et al., 2001).

Third, social software can be perceived as a mediator between groups of users and their respective positions. As a platform for exchange, a filter of information and knowledge, and as facilitator of organizational change, technology inevitably bears values and sides with certain perspectives that may reflect the organization only partially or privilege certain (powerful) individuals. Consumers can become fiercely critical of companies, management, or other employees (Kozinets and Handelman, 2004) and voice criticism even while generally advocating the brand they criticize (Muniz and O'Guinn, 2001). Social software may suddenly create opposing factions where they were previously hardly aware of each other. ICT may act as platform for the voices of consumers, users, or developers who loudly and explicitly vent what they could not say before or went unheard by management. Apart from information flows and employee motivation, such confrontation may require ethical deliberation from the strategist before and during the implementation of social software. Where does learning end in user communities and where does disruption for the company begin? What is the correct and appropriate level of respect towards emerging criticism, internal and external? How and when can social software be integrated into the work practice and become a balanced platform, guarantee equal access, or prevent uneven coverage of organizational events? The process by which social software co-evolves with organizations is strategically important, and the opportunities and limitations in managing and mediating co-evolution deserves more attention in future research.

Fourth, the development of social software as a new step in building virtual relations has given the trust issue a new edge. Fraudulent behavior, forgery and pretence have obviously not been spawned all of a sudden by the virtual world and social software. Questions concerning the original and the copy, not to mention the evaluation of informational goods that are the object of commercial transactions, have given rise to the problem of trust and have highlighted how crucial trust-building mechanisms are to the functioning of markets and communities since the beginning of time. But the development of virtual relations and social software has increased the need for new trust-building mechanism. What is at stake here is the entire range of mechanisms that will facilitate inter-personal and inter-organizational transactions, given the new conditions for knowledge transactions and exchanges: increasing specialization, increasingly asymmetrical distribution of information and assessment capabilities, ever greater anonymity among interlocutors and ever-more opportunities for forgery of identity. Clearly new methods need to be devised to "certify" the knowledge circulating through virtual relations within a context where inputs are no longer subject to control.

Contextualizing social software means studying social software as both the tool and mediator of organizational change triggered, facilitated, and aided by management. The word "contextualizing" implies a process, which is a process of construction, where users form networks, communicate across boundaries and exchange information that may alter their identities and work or question power relationships. That is why contextualizing social software may make power relationships transparent and brings forth ethical issues that researchers can analyze in the nascent structures of organizations. Actionable strategy research gives managers insights into accounts from other organizations about the demands put on them by very sophisticated or recalcitrant users, internal or external to the organization. On the one hand, internal users may undercut hierarchies by way of informal communications via social networks, and on the other hand managers scan Facebook entries before hiring. Issues pertaining to power relations and ethics run in several directions and call for research that makes these issues transparent and relates them to technology. Such research should also balance the perspectives between management and users, or internal and external stakeholders of the company.

**Co-existing with communities**

Social software plays an instrumental role in facilitating group work and bringing individuals together to form communities. Individuals gather around a shared purpose or attach to other members (Ren et al., 2007) resulting in communities who produce new technology (Sawhney and Prandelli, 2000) or celebrate certain forms of consumption (Muniz and O'Guinn, 2001). Online communities are organizations of their own rights that incorporate and govern their work (O'Mahony and Ferraro, 2007), enable joining and specialization of labor (von Krogh et al., 2003) and allow for firms to sponsor or regulate work (Shah, 2006; Bonaccorsi et al., 2006; West and O'Mahony, 2008; Capra et al., 2010). The value of social software-enabled communities for business seems obvious in terms of the knowledge they develop and conserve (Brown and Duguid, 2001). There are many types of online communities working on various purposes and breeding all sorts of interests and passions. Can companies emulate the best of the governance structures of online communities? And if yes, which type of community should serve as a template for learning? Or, more radically, will leadership need to be fundamentally recast in terms of open-ended notions of governance (Hess and Ostrom 2007)? The relationship between firms and online communities is not well understood in organization theory, where outlines became visible for such a relationship to communities located within the perimeters of the firm and operating face-to-face, nevertheless independent and "bottom-up" (Thompson, 2005). The attempt to gain influence in a community may amount to a struggle and a performance of influence with uncertain outcomes for the motivation jof users and their identification with the community's purpose. The study of leadership that bridges and connects firms and communities is an open field for management research.

Similarly, the discussion of community boundaries opens questions as to the purpose of a community, its membership base, and its dynamics. First, purpose may play a central role within the work practice and life context of the individuals who become members of the community (Muniz and O'Guinn, 2001). Communities bear and develop crucial knowledge in organizations (Brown and Duguid, 2001) and there is no reason to believe that communities where their users span organizational boundaries fall short on their ability to develop, protect, and share knowledge (e.g. Lerner and Tirole, 2002; Sawhney and Prandelli, 2000). Purpose and membership seem tightly linked not only for the posterior reason of defining what the community is about, but because individuals working on similar issues perceive the need to exchange and learn from their peers and mentors (Lave and Wenger, 1991). While this observation is general and pervades studies of collective action (Oliver, 1993; Ostrom, 1998) it translates into a series of questions regarding the use of social software whose technical implementation is approximately cost free using the Internet. If firms provide opportunities to interact individuals with similar interests are likely to pick up and exchange information. Importantly, the effects a community may have on the work practices within the organization are a direct consequence of the relatedness and bond that moved the individual to join the community in the first place.

Consider the fictitious example of a social software platform (such as LinkedIn or Xing) for the recruitment of management talent within a specific industry. Naturally, prospective talent will rush to become visible on the platform and so will recruiters. Given an open political and cultural context it becomes easy to see how labor market participants within that industry may join such an emerging community to discuss the firms' strategies, voice their ideas, and challenge each other's ideas. The platform may represent both a labor market opportunity as well as a branding and reputation challenge for participating and sponsoring firms. What policies should accompany the implementation of social software for such a platform? The issues include the eligibility to community membership and the authority to set boundaries, both internal and external. Co-existence with user communities means that authority for such policies is either shared with, or deferred to, members outside the organization, particularly if the company is only a marginal member.

Lastly, considering dynamic properties of community boundaries, the research issues become even more pronounced. There is particular value in search that bridges domains of knowledge (Poetz and Prügl, 2010; Laursen and Salter, 2006) and, thus, in community membership that expands in unpredicted ways. On the downside, reputation risks increase because of the nature of social software applications, we discussed above (Barwise and Meehan, 2010). Research on the dynamic properties of community boundaries is needed particularly regarding communities that extend beyond the boundaries of one company. Jarvenpaa and Lang (in this issue) suggest that boundaries are essential for the sustainability of the communities observed. This raises the important question about the link between company sponsorship and community boundaries, which has not been fully explored. The presence of a company in a user community may not only affect the members' motivations and work practices (Shah, 2006) but equally the membership dynamics and growth of the community.

## *Conclusion*

Social software challenges strategic thinking in important ways: the articles in this special issue show strategy practitioners meaningful ways to successfully deploy social software and strategy researchers which critical challenges deserve more attention. This introduction summarized the open research issues along three dimensions which we believe to be critically affected by the massive changes to everyday work in organizations, due to growing use and acceptance of social software within and across companies.

First, value creation and value appropriation can gain momentum through interaction with consumers and users inside and outside the firm. Javenpaa and Majchrzak (2010) speak of vigilant interaction to refer to collaboration with users that involves simultaneously sharing and protection of knowledge. The balance between sharing knowledge with consumers and users and protecting knowledge assets to appropriate value is subject to experimentation in practice and ongoing research in strategic management and organization theory. Users who had been careless about the rights to their contents and innovations (von Hippel, 1988) are becoming both more aware of available licenses for their intellectual property (ease of use of creative commens and Open Source licenses) and seemingly less sensitive about their privacy (using Twitter and location-based services such as foursquare or localuncle). A logic of co-creating strategy may extend the notion of emergent strategy to very active and loyal customers and users outside the company and create opportunities for strategists who understand and internalize the two perspectives of inside and outside the company.

Second, social software as a technology challenges not only competitive dynamics in industries but also the structure of organizations. With the increasing digitization of products and services interaction among consumers and users becomes easier and cheaper. My behavior as a customer of e-books depends on the device I use for reading it (mobile or at home etc.) and recommendations by friends and strangers. The competitive landscape is shaped by what Yoo, Henfridsson and Lyytinen (2010) call the layered modular architecture because the choice of a platform, to stay with the example of the e-book seller, is both a choice of hardware (the e-book reader) and of social network (for recommendations). Thus, social software may appear to be a tool of strategic choice. At the same time it is a mediator of relationships between the firm and users inside and outside the firm. Frey, Haag and Lütje (in this issue) make the point that empowering and restricting the user goes hand in hand with receiving substantive contributions by users. This balance is also a question of power relations and the design and implementation of social software needs to take into account that these relations can substantially alter or disrupt organizational processes (Leonardi, 2007).

Third, communities grow and build on social software applications that enable users and consumers to interact. Two choices impact strategic thinking: leadership and boundaries. To what extent should management lead a community and to what extent should strategists influence the extent of growth and influence of the community? Again, both decisions are constrained and enacted as part of a balancing act that makes the community possible in the first place. A logic of co-existence can guide strategic thinking when deciding about sponsoring a community by setting up social software infrastructures and sharing knowledge. The contribution by Denyer and colleagues in this issue alerts management to the pitfalls this can entail. The same logic may guide leadership that can be shared or distributed across community members who stand out independent and possibly outside of the firm.

A strategic approach to social software should start with the insight that empowering creative, independent individuals implies indeterminate and uncertain reactions and creations in support of or in opposition to the original thinking by management. New business opportunities abound and an experimental approach to strategy (McGrath, 2010) may be guided by first signposts erected by successful companies, who maintain long term relationships with their users. A number of them are described and analyzed in this special issue.